\documentclass[letterpaper, twocolumn, pra]{revtex4}
\usepackage{graphicx, amsmath, amssymb, bm, titlesec}

\begin{document}

\title{Efficient generation of propagation-invariant spatially-stationary partially coherent fields }

\author{Shaurya Aarav, Abhinandan Bhattacharjee, Harshawardhan Wanare, and Anand K. Jha}

\email{akjha9@gmail.com}

\affiliation{Department of Physics, Indian Institute of
Technology, Kanpur 208016, India}

\date{\today}

\begin{abstract}

We propose and demonstrate a novel method for generating
propagation-invariant spatially-stationary
fields in a controllable manner. Our method relies on producing incoherent mixtures of plane waves using planar primary sources that are spatially completely uncorrelated. The strengths of the individual plane waves in the mixture decide the exact functional form of the
generated coherence function. We use LEDs as the primary incoherent sources and experimentally demonstrate the effectiveness of our method by
generating several spatially-stationary fields, including a new type, which we refer to as the ``region-wise spatially-stationary field." 
We also experimentally demonstrate the propagation-invariance of these fields, which is an extremely interesting and useful property of such fields. Our work
should have important implications for applications that exploit the spatial coherence properties either in a transverse plane or in a propagation-invariant manner, such as correlation holography, wide-field OCT, and imaging through turbulence.

\end{abstract}

\maketitle

\section{INTRODUCTION}

Fields having partial spatial coherence have been
extensively studied in the past few decades \cite{mandel1995coherence, cai2014josaa, hyde2015jap} and have found a
wide range of applications including  wide-field optical coherence tomography (OCT) \cite{karamata2004optlett}, imaging through turbulence \cite{redding2012natphot}, optical communication
\cite{ricklin2002josaa, gu2010josaa}, particle trapping
\cite{zhao2009optexp, dong2012pra}, atomic optics
\cite{robb2007prl}, laser scanning \cite{kermisch1975josa}, plasma
instability suppression \cite{kato1984prl}, photographic noise
reduction \cite{belendez1993optcomm}, optical scattering
\cite{van2010prl}, and second harmonic generation
\cite{zubairy1987pra}. A spatially partially coherent field can be divided into two
categories: spatially stationary and spatially non-stationary. In analogy with the temporally-stationary fields, when the intensity of a field is independent of the spatial position and when the two-point spatial correlation function depends on the spatial positions only through their difference, the field is called spatially stationary, at least in the wide
sense \cite{takeda2005optexp, takeda2013optlett,
takeda2014or, turunen1991josaa, friberg1991pra, kowarz1995josaa, gori1987optcomm, naik2009optexp, naik2011optexp, ohtsuka1980optcomm}.  A spatially-stationary field has the unique property that its two-point correlation function is propagation invariant \cite{turunen1991josaa, ohtsuka1980optcomm}. Propagation-invariant, spatially stationary fields have several unique applications such as 3D coherence holography \cite{naik2009optexp} and photon correlation holography \cite{naik2011optexp}. If the field is not spatially-stationary, it is categorized as spatially non-stationary.

There are several different ways of producing spatially partially
coherent fields. While one of the earliest experiments used a laser and an acousto-optical cell \cite{ohtsuka1980optcomm}, later experiments utilized a laser and a rotating ground glass plate (RGGP) in order to produce fields with desired partial spatial
coherence \cite{chen2014pra, chen2014optexp, wang2014optexp,
ostrovsky2010optexp, wang2013optlett, wang2012apl,
wang2008optlett, takeda2005optexp, naik2009optexp}. More modern methods involve using a laser and either a spatial light
modulator (SLM) \cite{basu2014optexp,
ostrovsky2009optexp, ostrovsky2010optcomm, shirai2004josaa} or an RGGP in combination with an SLM to achieve the purpose
\cite{liang2014optlett, chen2014optexp2, naik2011optexp}. As far as propagation-invariant spatially-stationary partially coherent fields are concerned, to the best of our knowledge,
there have been only two experimental studies so far. In the first experiment the field was generated using a laser and an acousto-optic cell \cite{ohtsuka1980optcomm} and in the second experiment the generation was done using a laser and an RGGP \cite{turunen1991josaa}. Nevertheless, both these techniques have demonstrated generation of only those cross-spectral density functions that can be represented as Fourier transforms of circularly-symmetric functions.

\begin{figure}[b!]
\includegraphics{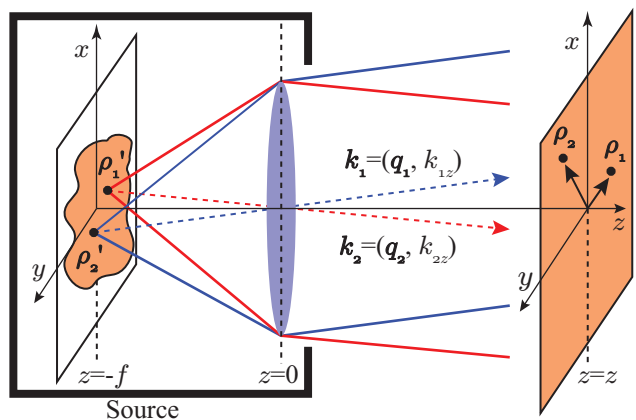}
\caption{(color online) Schematic illustration of how a propagation-invariant spatially-stationary field can be generated using a spatially completely-uncorrelated primary source.}
\label{fig1}
\end{figure}

Thus, all the existing experimental techniques for producing spatially-stationary partially coherent fields and most techniques for producing spatially non-stationary partially coherent fields use a laser as the primary source, which, to begin with, is spatially a completely correlated source. One then tries to make the field emanating from such a source spatially partially
coherent by introducing randomness in the field-path by using either an acousto-optic cell \cite{ohtsuka1980optcomm}, or an RGGP \cite{chen2014pra, chen2014optexp, wang2014optexp,
ostrovsky2010optexp, wang2013optlett, wang2012apl,
wang2008optlett, takeda2005optexp, naik2009optexp} or an SLM \cite{basu2014optexp,
ostrovsky2009optexp, ostrovsky2010optcomm, shirai2004josaa,liang2014optlett, chen2014optexp2, naik2011optexp}. On the other hand, in this article, we propose a
technique that uses a primary source that is spatially completely
uncorrelated, and we demonstrate
generation of very high-quality propagation-invariant spatially stationary fields, without having to introduce any additional randomness. Furthermore, we show that our technique can produce any propagation-invariant spatially-stationary cross-spectral density function and not just the ones that are Fourier transforms of circularly-symmetric functions \cite{turunen1991josaa, ohtsuka1980optcomm}. 

\begin{figure*}[t!]
\includegraphics{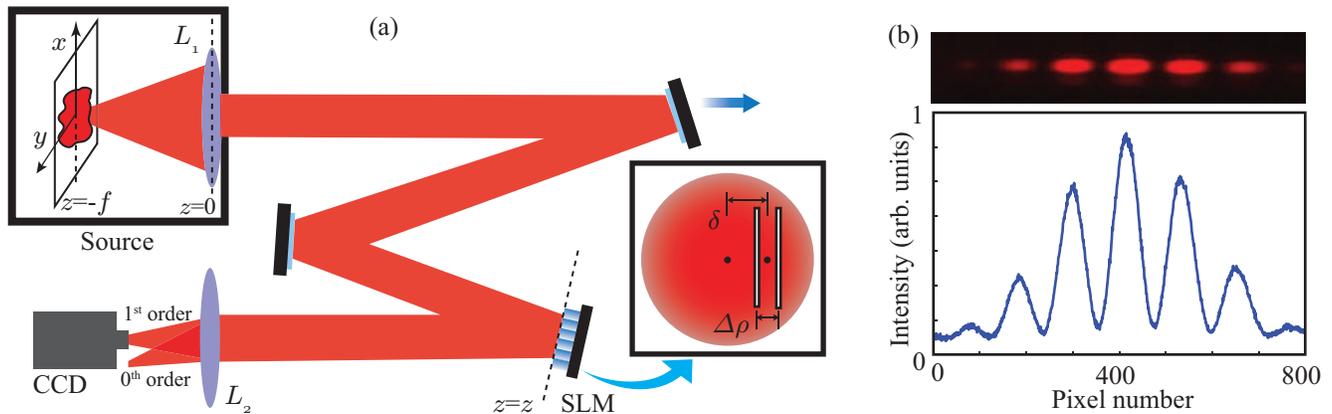}
\caption{(color online) (a) Schematic diagram of the experimental
setup. A planar, spatially incoherent primary source is placed at the back focal plane of lens $L_1$. The cross-spectral density of the field produced by the sourse is measured using the Spatial Light Modulator (SLM). The propagation length $z$ is the distance between
the lens $L_1$ and the SLM, and the CCD-camera is placed at the focal plane of lens $L_2$. (b) A representative experimental interference pattern produced by the double-slit simulated on the SLM, and the associated one-dimensional plot.
 }\label{fig2}
\end{figure*}

\section{THEORY: PROPAGATION-INVARIANT SPATIALLY-STATIONARY PARTIALLY COHERENT FIELDS}

Let us consider the situation shown in Fig.~\ref{fig1}. A planar, monochromatic, spatially completely incoherent primary source is kept at the back focal plane $z=-f$ of a lens kept at $z=0$. The planar primary along with the lens constitute our source of spatially partially coherent fields. We represent the field radiating out from spatial location $\bm\rho'$ at $z$ by $V_s(\bm\rho', z)$. Since our primary source is spatially completely incoherent, the fields $V_s(\bm\rho'_1, -f)$ and $V_s(\bm\rho'_2, -f)$ radiating out from $\bm\rho'_1$ and $\bm\rho'_2$, respectively, at $z=-f$ are completely uncorrelated, that is,
\begin{align}
\langle V_s^*(\bm\rho'_1, -f)V_s(\bm\rho'_2, -f)\rangle_e=I_s(\bm\rho'_1, -f)\delta(\bm\rho'_1-\bm\rho'_2).\label{position-correlation}
\end{align}
Here $I_s(\bm\rho'_1, -f)$ is the intensity of the primary source at $z=-f$. We note that no realistic primary source can truly have a position correlation given by Eq.~(\ref{position-correlation}), which requires that the spatial coherence length be zero. The smallest spatial coherence length that can be associated with a primary source is of the order of the wavelength $\lambda$ of the source, and only a black body emitter can be idealized as such a source \cite{carter1977josa}. Nevertheless, for a millimeter-size source at optical wavelengths, the position correlations of the order of $\lambda$ can very well be approximated by Eq.~(\ref{position-correlation}). In our experiments, we use LEDs as our primary incoherent sources, which are considered spatially completely incoherent in the sense that their position correlations are approximated by the form given in Eq.~(\ref{position-correlation}) \cite{tziraki2000apb, pires2010optlett}. 

Thus, for our primary source whose position correlation is represented by Eq.~(\ref{position-correlation}), every point on the source is radiating out as an independent point source and since each of these points is kept at the back focal plane of a converging lens, the field $V_s(\bm\rho'_1, -f)$ radiating out from $\bm\rho'_1$ gets transformed into a plane wave with amplitude $a(\bm{q_1})$ by the lens, where $\bm{q_1}$ represents the transverse wave-vector associated with the plane wave \cite{born1999, goodman2005}. Here, we are assuming that the aperture-size of the lens is infinite. This turns out to be a very good approximation for our purposes in this section and the next; the effects due to a finite aperture-size lens is discussed and demonstrated in Sec.~IV. The lens, therefore, transforms the non-correlation of the planar source in the position basis to non-correlation in the transverse wave-vector basis. The correlations between different transverse wave-vectors are quantified using the angular correlation function $\mathcal{A}(\bm{q_1}, \bm{q_2})$. It is defined as $\mathcal{A}(\bm{q_1}, \bm{q_2}) \equiv \langle a^*(\bm{q_1})a(\bm{q_2})\rangle_e$ where  $\langle\cdots\rangle_e$ represents the ensemble average. The angular correlation function of our partially coherent source is the angular correlation function $\mathcal{A}(\bm{q_1}, \bm{q_2})$ at the exit face of the lens, that is, at $z=0$, and is thus given by 
\begin{align}
\mathcal{A}(\bm{q_1}, \bm{q_2}) \equiv \langle
a^*(\bm{q_1})a(\bm{q_2})\rangle_e=I_s(\bm{q_1})\delta(\bm{q_1}-\bm{q_2}).\label{angular-correlation-fn}
\end{align}   
Here $I_s(\bm{q_1})$ is the spectral density of the field; it has the same functional form as that of the source intensity. As we show below, this form of the angular correlation function is \textit{the} requirement for the partially-coherent field coming out of a source to be spatially-stationary and propagation-invariant. 

We next derive the cross-spectral density function at $z=z$  produced by our source. As worked out in Section 5.6 of Ref.~\cite{mandel1995coherence}, if the plane-wave amplitude at $z=0$ is $a(\bm{q_1})$ then the field $V(\bm\rho_1, z)$ at $z=z$ within the paraxial approximation is given by
\begin{align}
V(\bm\rho_1, z)=e^{ik_0z}\iint_{-\infty}^{\infty}
a(\bm{q_1})e^{i\bm{q_1}.\bm\rho_1}e^{-i\tfrac{q_1^2 z}{2k_0}}d\bm{q_1}.
\end{align}
Here, we have used the fact that $\bm{r_1}\equiv(\bm{\rho_1}, z)$,
$\bm{k_1}\equiv(\bm{q_1}, k_{1z})$, and $k_{1z}\approx k_1-q_1^2/(2 k_1)$ with $q_1=|\bm{q_1}|$ and $k_{1}=|\bm{k_1}|=k_0=\omega_0/c$, where $\omega_0$ is the frequency of the field. The cross-spectral density function $W(\bm\rho_1, \bm\rho_2, z)\equiv \langle V^*(\bm\rho_1, z) V(\bm\rho_2, z) \rangle_e$ at $z=z$ is therefore 
\begin{align}
W({\bm{\rho}_1}, {\bm{\rho}_2}, z)= &
\iint_{-\infty}^{\infty} \mathcal{A}(\bm{q_1}, \bm{q_2})
\notag\\ \times & e^{-i\bm{q_1}.\bm\rho_1+i\bm{q_2}.\bm\rho_2}
e^{-i\tfrac{(q_1^2-q_2^2)z}{2k_0}}d\bm{q_1} d\bm{q_2}.\label{cs-density}
\end{align}
Equation (\ref{cs-density}) governs how spatial correlations of the field, as represented by the cross-spectral density function, change upon propagation in the region $z>0$ after the lens. Substituting the form of the angular correlation function from Eq.~(\ref{angular-correlation-fn}) into Eq.~(\ref{cs-density}), we obtain
\begin{align}
W({\bm{\rho}_1}, {\bm{\rho}_2}, z)= W(\bm\Delta\rho, z)= 
\int_{-\infty}^{\infty} I_s(\bm{q})
 e^{-i\bm{q.\Delta\rho}}d\bm{q},\label{cs-density2}
\end{align}
where $\bm\Delta\rho=\bm{\rho_1}-\bm{\rho_2}$. The intensity $I(\bm\rho, z)$ corresponding to the above cross-spectral density function is
\begin{align}
I(\bm\rho, z)=W(\bm{\rho}, \bm{\rho}, z)= 
\int_{-\infty}^{\infty} I_s(\bm{q})d\bm{q}=K,\label{intensity}
\end{align}
where $K$ is a constant. We find that the cross-spectral density function $W(\bm\Delta\rho, z)$ in Eq.~(\ref{cs-density2}) is in the coherent-mode representation, with the plane waves being the coherent modes. In other words, our source produces a field that is an incoherent mixture of plane-wave modes. As a result, the generated field has the following properties: (1) \textit{The field is propagation invariant}---This is because the cross-spectral density function as well as the intensity is independent of $z$.  (2) \textit{The field is spatially stationary at a given $z$, at least in the wide sense.}---This can be verified by noting that the intensity $I(\bm\rho, z)$ does not depend on $\bm\rho$ and the cross-spectral density function depends on $\Delta\bm\rho$ only. (3) \textit{The cross-spectral density function $W(\bm\Delta\rho, z)$ of the field is the Fourier transform of its spectral density $I_s(\bm{q})$}---this is the spatial analog of the Wiener-Khintchine theorem for temporally-stationary
fields (see Section 2.4 of \cite{mandel1995coherence}). Moreover, since the spectral density has the same functional form as the intensity of the primary source, the cross-spectral density function of the field is the Fourier transform of the intensity profile of the primary source. We note that in our technique there is no restriction on the form of the intensity function $I_s(\bm{q_1})$ that the primary incoherent source can have. The primary source can be continuous or having a finite size or even in the form of a collection of points. As a result, using our technique, one can produce any custom-designed, spatially-stationary propagation-invariant partially coherent field and not just the ones that are Fourier transforms of circularly-symmetric functions \cite{turunen1991josaa, ohtsuka1980optcomm}.

\begin{figure}[t!]
\includegraphics{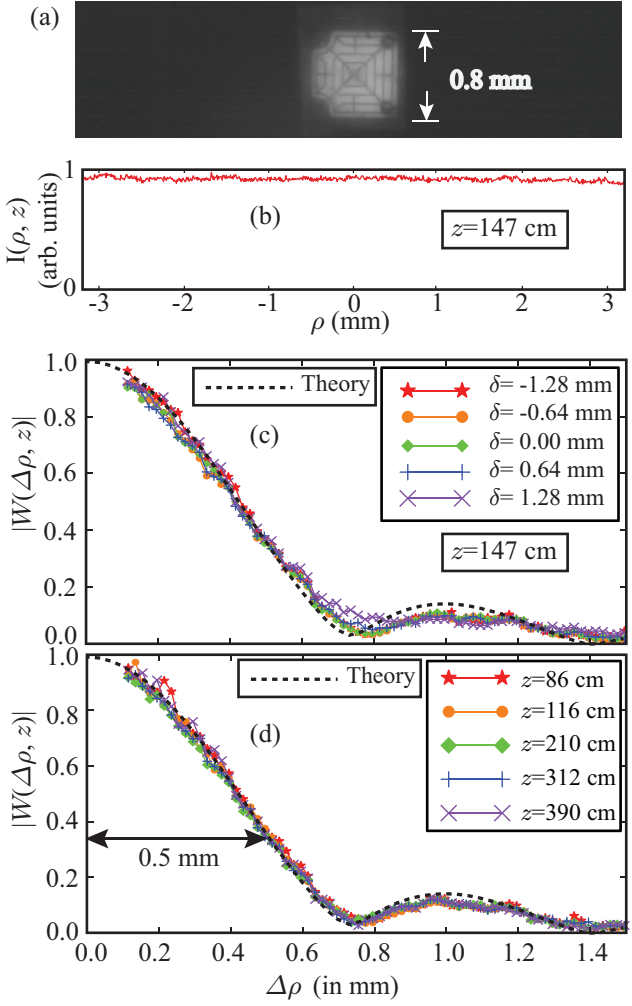}
 \caption{(color online) (a) The CCD-camera image of the LED. (b) Plot of intensity $I(\bm\rho, z)$ as a function of $\bm\rho$ at $z=147$ cm. (c)  Plots of $|W(\bm\Delta\rho, z)|$ as a function of $\bm\Delta\rho$ at $z=147$ cm for various values of the offset parameter $\delta$. (d) Plots of $|W(\bm\Delta\rho, z)|$ as a function of $\bm\Delta\rho$ for various values of $z$. In the above figures, the black dashed curves represent the theoretical prediction based on Eq.~(\ref{cs-density2}).}\label{fig3}
\end{figure}

\begin{figure}[t!]
\includegraphics{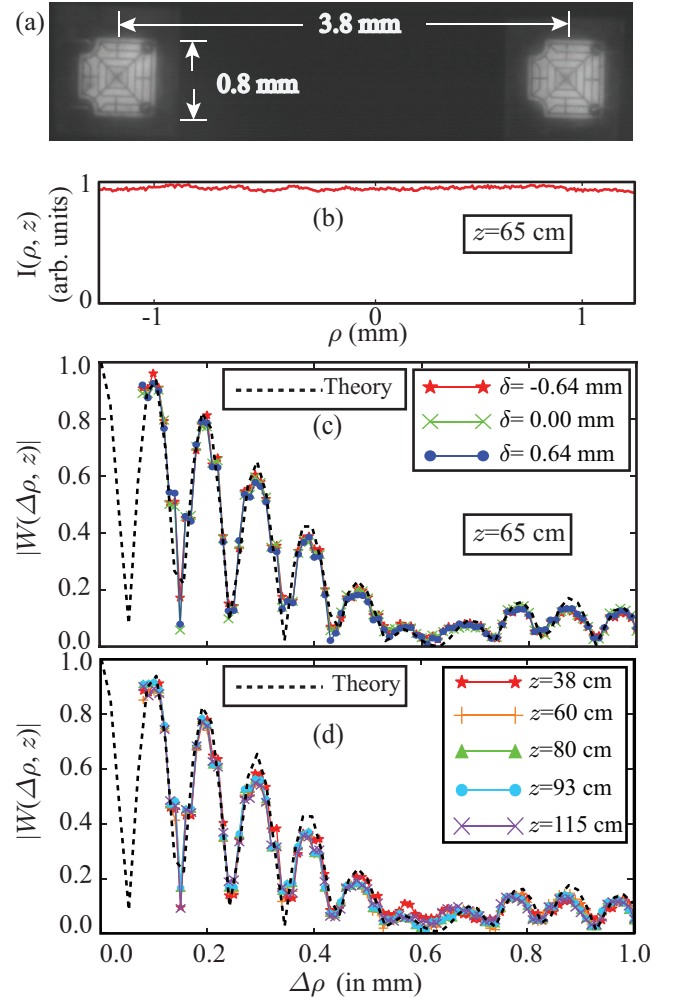}
\caption{(color online).(a) The CCD-camera image of the LED. (b) Plot of intensity $I(\bm\rho, z)$ as a function of $\bm\rho$ at $z=65$ cm. (c) Plots of $|W(\bm\Delta\rho, z)|$ as a function of $\bm\Delta\rho$ at $z=65$ cm for various values of the offset parameter $\delta$. (d) Plots of $|W(\bm\Delta\rho, z)|$ as a function of $\bm\Delta\rho$ for various values of $z$. In the above figures, the black dashed curves represent the theoretical prediction based on Eq.~(\ref{cs-density2}).}\label{fig4}
\end{figure}

\section{EXPERIMENTAL DEMONSTRATIONS}

Fig.~\ref{fig2}(a) shows the schematic of our experimental setup. Our primary source is a commercially available 9 W planar LED bulb. We use an interference filter centered at 632.8 nm having a wavelength-bandwidth of 10 nm. The LED bulb consists of 9 separate LEDs arranged in a 3$\times$3 grid. We take the individual LEDs to be spatially completely incoherent \cite{tziraki2000apb, pires2010optlett} in the sense that their spatial-correlation function can be approximated by Eq.~(\ref{position-correlation}). The individual LEDs are of dimensions $0.8\times 0.8$ mm and the separation between two nearest LEDs is $1.9$ mm. We let the field produced by our source at $z=0$ propagate to $z=z$ and then measure the cross-spectral density function using a Young's double-slit pattern simulated on an SLM kept at $z=z$ \cite{savage2009natphot,gruneisen2008ao,jhonson1993ieee}, with the separation between the slits being $\bm\Delta\rho$. The distance between the center of the field and the center of the double-slit is the offset parameter $\delta$. We record the resulting interference fringe pattern by keeping a CCD camera at the focal plane of lens $L_2$ and then capturing only the first diffraction order due to the SLM. We note that since the two simulated slits are exactly the same and since the field is uniform in intensity, the magnitude $|W(\bm\Delta\rho, z)|$ of the cross-spectral density function is the visibility of interference fringes. Therefore, by measuring the interference visibility as a function of the slit separation $\bm\Delta\rho$, we directly measure $|W(\bm\Delta\rho, z)|$ as a function of  $\bm\Delta\rho$. We further note that any pattern simulated on an SLM is seen by only one polarization component of an incoming field \cite{jhonson1993ieee}, and it is only this component that contributes at the first diffraction order. The other polarization component, if present, simply ends up at the zeroth diffraction order. Since our measurements are made only at the first diffraction order, only one polarization component gets measured and therefore scalar theory of Sec.~II should be sufficient to describe the present experiments.

A typical interference pattern observed using the CCD camera and the associated one-dimensional section of the intensity pattern are shown in Fig.~\ref{fig2}(b). Figure \ref{fig3}(a) is the image of the central LED of our bulb. First of all, we make measurements with this being our primary source. The focal length $f$ of lens $L_1$ is $75$ cm. Figure \ref{fig3}(b) shows the plot of the intensity at $z=147$ cm and Fig.~\ref{fig3}(c) shows plots of $|W(\bm\Delta\rho, z)|$  at $z=147$ cm as a function of $\bm\Delta\rho$ for several offset values $\delta$. These results verify that the generated field is spatially stationary. Figure \ref{fig3}(d) shows plots of $|W(\bm\Delta\rho, z)|$ as a function of $\bm\Delta\rho$ for various propagation distances up to 3.9 m. There is little variation between the different plots. This proves that the cross-spectral density function of the generated field is propagation invariant at least up to a distance of $3.9$ meters. We note that the transverse coherence length of the field, which we define to be the value of $\bm\Delta\rho$ at which $|W(\bm\Delta\rho, z)|$ drops down to $1/e$, is about 0.5 mm and remains propagation invariant. This is in contrast to the field produced by a bare primary source of the same shape and size as that of the source in Fig.~\ref{fig3}(a), in which case the transverse coherence length, following the conventional van-Cittert Zernike theorem, increases by about $5$ times after propagating for $4$ meters.

Next, we make measurements with our primary source containing two spatially separated LEDs. The image of the primary source is shown in Fig.~\ref{fig4}(a). Figure \ref{fig4}(b) shows the plot of the intensity at $z=65$ cm. Figure \ref{fig4}(c) shows plots of $|W(\bm\Delta\rho, z)|$ as a function of $\bm\Delta\rho$ for various values of the offset parameter $\delta$ at $z=65$ cm, and Fig.~\ref{fig4}(d) shows plots of $|W(\bm\Delta\rho, z)|$ as a function of $\bm\Delta\rho$ at various $z$. These results again demonstrate spatial stationarity and propagation invariance. It is interesting to note that the cross-spectral density function in this case is in the form of a fringe pattern which is nothing but the Fourier transform of our source shown in Fig.~\ref{fig4}(a). 

Using Eq.~(\ref{cs-density2}) and the image of our primary sources shown in Figs.~\ref{fig3}(a) and \ref{fig4}(a), we also calculate the theoretical cross-spectral density functions and plot them along with the experimental results in Figs.~\ref{fig3} and ~\ref{fig4}. Our reported experimental results match very well with the theoretical predictions, demonstrating the accuracy and effectiveness with which a custom-designed, spatially-stationary propagation-invariant cross-spectral density function can be generated using our method. In order to produce a field with a given cross-spectral density function one simply needs to construct a primary source with an intensity distribution that is the inverse Fourier transform of the desired cross-spectral density function.

\begin{figure}[t!]
\includegraphics{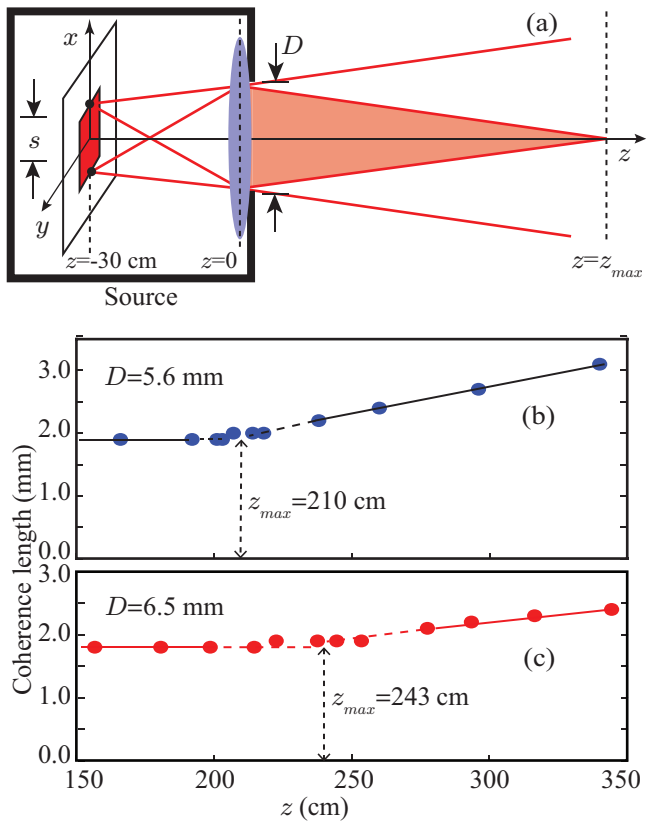}
\caption{(color online) (a) Diagram illustrating how the aperture-size $D$ of the lens and the spatial width $s$ of the primary source fixes $z_{\rm max}$. (b) and (c) show plots of the transverse coherence length as a function of $z$ for $D=5.6$ mm and $D=6.5$ mm, respectively.}\label{fig5}
\end{figure}

\begin{figure}
\includegraphics{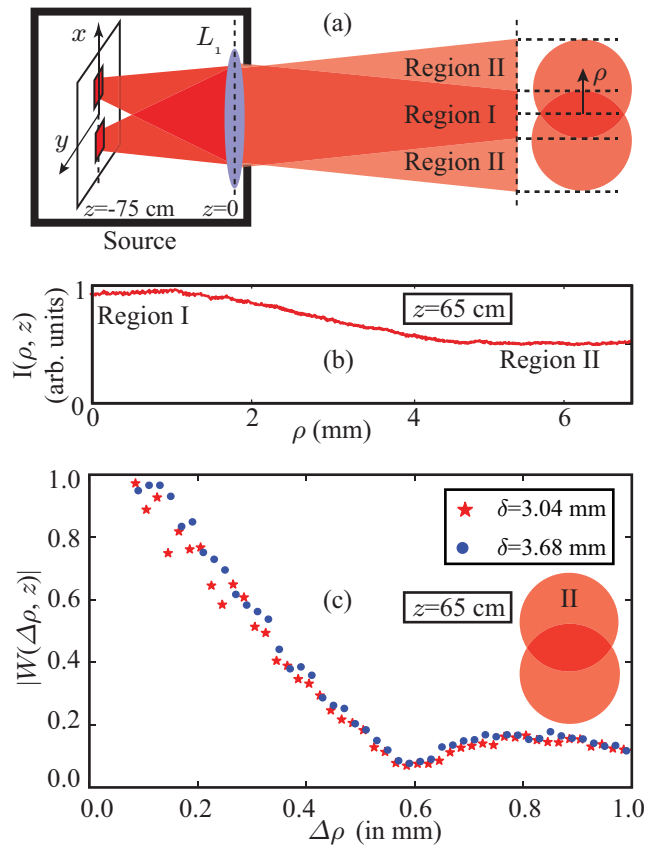}
\caption{(color online) (a) Diagram illustrating the generation of region-wise spatially stationary fields. (b) Plot of intensity $I(\bm\rho, z)$ as a function of $\bm\rho$ at $z=65$ cm. (c) Plots of $|W(\bm\Delta\rho, z)|$ as a function of $\bm\Delta\rho$ at $z=65$ cm for various values of the offset parameter $\delta$.}\label{fig6}
\end{figure}

\section{EFFECTS DUE TO A FINITE-SIZE LENS}

The theoretical modeling presented so far assumes that the lens that constitutes our partially coherent source has an infinite aperture-size. However, in a realistic experimental situation the aperture size of a lens is finite, and in our case it is of the order of an inch. As discussed in Ref.~\cite{turunen1991josaa}, and as illustrated in Fig.~\ref{fig5}(a), the finite aperture-size of the lens restricts the propagation invariance properties to distance $z_{\rm max}$, given by $z_{\rm max}=Df/s$, where $D$ is the aperture size of the lens, $f$ is the focal
length and $s$ is the size of the primary source. In order to experimentally demonstrate $z_{\rm max}$, we used the LED source shown in Fig.~\ref{fig3}(a) with an $f=30$ cm lens. Figures \ref{fig5}(b) and \ref{fig5}(c) show how the transverse coherence length changes as a function of $z$ for two different values of the aperture-size $D$. As the aperture-size becomes bigger $z_{\rm max}$ gets larger. Nevertheless, even with realistic aperture sizes, one can easily achieve a $z_{\rm max}$ of up to tens of meters. 

Although the finite aperture-size of the lens may seem to only have the restricting effect on $z_{\rm max}$, it can in fact lead to restructuring of spatial correlations in a way that can have its own set of advantages. We now report such a restructuring effect when the primary source is in the form of two spatially separated LEDs, as shown in Fig.~\ref{fig4}(a). As illustrated in Fig.~\ref{fig6}(a), the propagation-invariant field generated due to such a primary source has two distinct regions over which spatial stationarity is observed. Region-I receives plane wave contributions from both the LEDs while Region-II receives the contributions from a single LED only. This leads to the two regions having two distinct spatially-stationary propagation-invariant cross-spectral density functions. We term such fields as ``region-wise spatially stationary fields." Figure \ref{fig6}(c) shows the plots of $|W(\bm\Delta\rho, z)|$ as a function of $\bm\Delta\rho$ for various values of the offset parameter $\delta$ in Region-II. These results demonstrate the spatial stationarity in Region-II. The spatial stationarity of Region-I is already shown in Fig.~\ref{fig4}(c). Therefore, the finite aperture-size of the lens offers an advantage in creating region-wise spatially stationary fields.

\section{CONCLUSIONS}

In conclusion, in this article, we have proposed and demonstrated a
novel method for generating custom-designed,
propagation-invariant, spatially-stationary fields. Our method can be used for generating any spatially-stationary cross-spectral density function as long as it has a coherent-mode-representation in the plane-wave basis.  Our experimental technique is based on using a spatially uncorrelated primary source and does not require introduction of any additional randomness, as is required by most other conventional methods. We have experimentally demonstrated the effectiveness of this technique by generating different spatially-stationary fields, including a new,
region-wise spatially-stationary field. We have also  demonstrated propagation invariance up to a few meters for several of these
spatially-stationary fields. The high-efficiency and control inherent in our technique can have important practical implications for several applications.  The propagation-invariant spatially-stationary fields are already a necessity for applications such as correlation holography \cite{naik2009optexp, naik2011optexp}. We belive that such fields can be an enabler for the 3D version of imaging through turbulence \cite{redding2012natphot} and wide-field OCT \cite{karamata2004optlett}. Moreover, the region-wise spatially stationary fields could provide unique benefits when the feature-sizes are spatially non-uniform.

\section*{ACKNOWLEDGMENTS}

We acknowledge financial support through an initiation
grant no. IITK /PHY /20130008 from Indian Institute
of Technology (IIT) Kanpur, India and through the research
grant no. EMR/2015/001931 from the Science
and Engineering Research Board (SERB), Department
of Science and Technology, Government of India.


\end{document}